\documentclass[aip,jcp,reprint,showpacs,floatfix,groupedaddress]{revtex4-1}
\usepackage[dvips]{graphics}
\usepackage{graphicx}
\usepackage[utf8]{inputenc}
\usepackage{amssymb}
\usepackage{epstopdf}
\usepackage{bm}

\def\s#1{_{\rm #1} }

\def\e{ {\rm e } }
\def\bea{\begin{eqnarray}}
\def\eea{\end{eqnarray}}
\def \be{\begin{equation}}
\def \ee{\end{equation}}

\begin{document}
\title{Theory of photoferroelectric response in SmC* liquids}
\author{Milo\v{s} Kne\v{z}evi\'{c}}
\author{Mark Warner}
\email{mw141@cam.ac.uk}
\affiliation{Cavendish Laboratory, University of Cambridge, JJ Thomson Avenue,
Cambridge CB3 0HE, United Kingdom}
\date{\today}

\begin{abstract}
We are concerned with the modification of liquid crystalline and polar order
in SmC* liquids by illumination. In particular we show that non-uniformity due to absorption and
also dynamics, can be complex. The variation of polarization with temperature, while
illuminated, is modified from that assuming uniformity.
Apparent changes of polarization with illumination will be shown to be
underestimated due to non-uniformity. The dynamics is shown to depend on
propagating fronts of photo-conversion penetrating the sample.
\end{abstract}

\maketitle

\section{Introduction}

\begin{figure}
  \includegraphics[width=7.8cm]{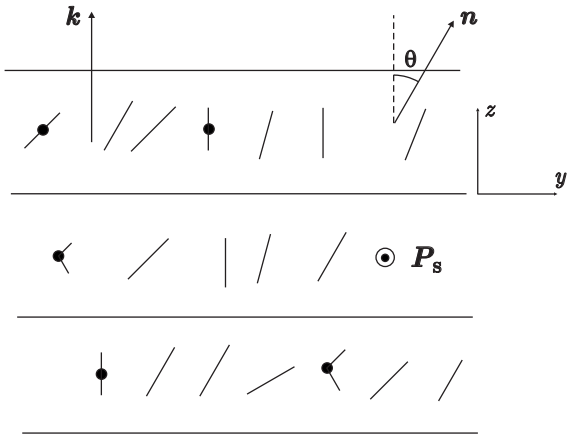}
  \caption{A smectic C* liquid host. Guest dye molecules with a photo-active
center indicated by a dot are also shown, some in their linear (\textit{trans}) ground state,
others bent, in their excited (\textit{cis}) state.}
  \label{fig1}
\end{figure}

Smectic C liquids possess an underlying nematic, or orientational ordering of
their molecular rods.
Additionally the rods have some ordering into layers, with layer normal
$\bm{k}$, and their nematic director $\bm{n}$ is tilted by an angle $\theta$ with respect to $\bm{k}$
(Fig.~\ref{fig1}).

When the rods are chiral we have a SmC* liquid where the plane of $\bm{k}$ and
$\bm{n}$ is no longer a mirror plane. As first recognized by Meyer~\cite{meyer:75}, a SmC* admits of
a spontaneous polarization along $\bm{k} \wedge \bm{n}$, that is in the layers and perpendicular to
$\bm{n}$ and $\bm{k}$.
Indeed Pikin and Indenbom~\cite{pikin:78} pointed out that $(\bm{n} \cdot
\bm{k})(\bm{n} \wedge \bm{k})$ is the appropriate order parameter and that the polarization, $\bm{P_s}$, can be
expressed as
\be
\bm{P\s{s}} = \varepsilon_0 \chi_0 c^{*} (\bm{n} \cdot \bm{k})(\bm{n} \wedge
\bm{k}), \quad P\s{s} = \varepsilon_0 \chi_0 c^{*} \theta,
\ee
where $\varepsilon_0$ is the standard vacuum permittivity, $\chi_0$ is the generalized susceptibility of a corresponding racemate in the direction perpendicular to $\bm{n}$ and $\bm{k}$, and $c^{*}$ is the so-called bilinear coupling coefficient. The latter expression for $P\s{s}$ holds for small tilt (see Lagerwall's book~\cite{lagerwallbook:99} for a review; we adopt his convention for the
coupling $c^{*}$, with an asterisk indicating its chiral origin). The
polarization is termed ``pseudo proper ferroelectricity'' since the primary order
parameter is not the polarization but derives from the tilt which in turn
drives the polarization.

On increasing temperature, the SmC* phase undergoes a second-order phase
transition to the SmA* phase. The tilt angle $\theta$ of the SmC* phase vanishes near the
critical temperature $T\s{AC}$, and consequentially $P\s{s}$ too. It is expected that this
transition belongs to $d=3$, $n=2$ universality class~\cite{deGennesprostbook:93}.

It has been long known that the order can be changed by irradiation of SmC*
containing photo-active rods as guests~\cite{ikeda:93n,ikeda:93f,coles:93l,coles:93f,coles:94l}.
By absorbing a photon dye molecules can make transitions from the ground (\textit{trans}) state
to the excited bend-shaped (\textit{cis}) state. The molecules in the \textit{cis} state either disrupt
the SmC* ordering to change $\theta$, at a given $T$, or weaken the coupling $c^{*}$. Both changes are a route to
changing $P\s{s}$, hence the name photoferroelectric effect. Langhoff and Giesselmann were concerned with
attributing photo changes to primary ($c^{*}$) or secondary ($\theta$) influences~\cite{giesselmann:03}.

We are concerned with the character of the light absorption. Since there has to
be a finite change in the ground state population of absorbers, then the absorption can not be
linear, that is it must be non-Beer in form. Moreover the detailed dynamics of $P\s{s}$ and $\theta$
observed after the start of illumination indicates nonlinear effects. In any event, polarization is
likely to vary pronouncedly with depth into the sample, especially away from $T\s{AC}$ and at not too high
intensities of incident light. We show how to accommodate this non-uniformity in modeling the photo response.
We suggest experiments to correlate $P\s{s}$ with mechanisms of absorption.

We first sketch in Sect.~\ref{sect:abs} the forms of light absorption leading
to non-uniform polarization. We then present a simplest model of how to translate
the dark state polarization into the light state, and superimpose such results to model
experiment. In Sect.~\ref{sect:dyn} we introduce dynamics to construct the
analogous time-dependent forms of the measured polarization. Switch-on dynamics depends on
propagating fronts of photo-conversion penetrating the sample, while switch-off dynamics is
essentially a relaxation type dynamics.

\section{Light absorption and non-uniformity of
photo-conversion}\label{sect:abs}

\subsection{Light absorption}

\begin{figure}
  \includegraphics[width=8cm]{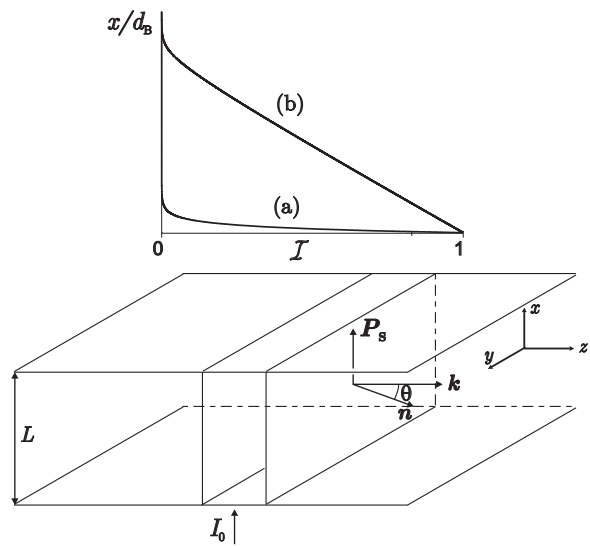}
  \caption{A SmC* sample between two electrodes illuminated from below ($x=0$)
with light of intensity
$I_0$. The decay of scaled intensity $\mathcal{I}(x) = I(x)/I_0$ is shown above: (a) exponential (Beer) decay,
$\alpha=0$; (b) deep penetration (non-Beer), in this example $\alpha=30$.}
  \label{fig2}
\end{figure}

The intensity, $I$, of light varies with depth, $x$, due to absorption by
\textit{trans}-species of dye molecules
\be
\frac{\partial I}{\partial x} = - \gamma \Gamma I(x,t) n\s{t}(x,t),
\label{intensity}
\ee
where $\Gamma$ is a rate coefficient for photon absorption and $\gamma = \hbar
\omega n\s{d}$ subsumes the energy $\hbar \omega$ of each
absorption of a photon from the beam and the absolute
number density of chromophores, $n\s{d}$. The number fraction of these
chromophores in their \textit{trans} state is $n\s{t}$; clearly $n\s{t} + n\s{c} = 1$, where
$n\s{c}$ is the number fraction of chromophores in the \textit{cis} state. To
simplify dynamics we neglected absorption by the \textit{cis}-species. This can be justified by
the fact that most experiments performed so far used
light wavelengths in the vicinity of the \textit{trans} absorption maximum.
We normalize $I(x,t)$ by the incident intensity $I_0$ to give an
$\mathcal{I}(x,t) = I(x,t)/I_0$.
The combination $\gamma \Gamma$ will be written $1/d\s{B}$ with $d\s{B}$ the
Beer length. Eq.~(\ref{intensity}) can be rewritten as
\be
\frac{\partial \mathcal{I}}{\partial x} = - \frac{n\s{t}}{d\s{B}} \mathcal{I}.
\label{intensityr}
\ee
If the \textit{trans}-population of absorbers is assumed to change little,
$n\s{t} (x,t) \simeq 1$, then one has Beer attenuation $\mathcal{I} = \e^{-x/d\s{B}}$. Only if the sample
thickness, $L$, is much less than $d\s{B}$ do we obtain a roughly uniform response, though in the Beer
limit we assume essentially no response, $n\s{t} \simeq 1$. Eq.~(\ref{intensityr}) can only be
closed if we know the \textit{trans}-population at ($x$,$t$). We have
\be
\frac{\partial n\s{t}}{\partial t} = - \eta \Gamma I(x,t) n\s{t} +
\frac{n\s{c}(x,t)}{\tau},
\label{transnumber}
\ee
where changes in $n\s{t}$ are due to conversions, with quantum efficiency
$\eta$ per photon absorption of \textit{trans}-\textit{cis} transition, and thermal back reaction from
\textit{cis}-population at a rate $1/\tau$, with $\tau$ the \textit{cis} lifetime. In the steady state, $\partial
n\s{t}/ \partial t = 0$, we have
\be
n\s{t}(x) = \frac{1}{1 + \alpha \mathcal{I}}.
\label{transeq}
\ee
We adopt the convention: where the $t$ argument is absent, we are denoting the
steady state values of $n\s{t}$ and $\mathcal{I}$. The combination $\alpha = \eta \tau \Gamma I_0
\equiv I_0/I\s{t}$ is a measure of how intense the incident beam is compared with a material constant
$I\s{t}$ for the \textit{trans} species. The parameter $\alpha$ is a balance between the forward rate, $\eta
\Gamma I_0$, and the back rate, $1/\tau$.
The solution of Eq.~(\ref{intensityr}) in the equilibrium case can be obtained by using
relation~(\ref{transeq})
\be
\ln [\mathcal{I}(x)] + \alpha [\mathcal{I}(x) - 1] = - \frac{x}{d\s{B}}.
\label{eqsolution}
\ee
For $\alpha =0$, we have the usual exponential form of Beers law.
For large $\alpha$ we have linear rather than exponential penetration
\be
\mathcal{I}(x) \simeq 1 - \frac{x}{\alpha d\s{B}}
\label{connonlin}
\ee
at least over depths up to $x \sim \alpha d\s{B}$ whereupon $\mathcal{I}$ is
small and the $\ln (\mathcal{I})$ again prevails to give a finally exponential penetration (see curves (a) and
(b) in Fig.~\ref{fig2} for examples of both profiles).
Thus, depletion of the absorber population at high intensities can give very
deep penetration to $x \gg d\s{B}$.
Non-Beer absorption was first explored for
dyes in liquid crystals (nematics) by Statman and Janossi~\cite{statman:03} and by Corbett and
Warner~\cite{corbettdyn:03,corbettdyn:02}.
The determinant, $\alpha$, of whether the absorption is non-linear depends on
temperature $T$ since the \textit{cis}-\textit{trans} decay is activated. The excited state thermal life
time takes an Arrhenius form $\tau = \tau_0 \e^{\kappa/T}$. We shall require $n\s{c}(x,\alpha)$ as a function
of $T$ which enters through $\alpha(T)$.

\subsection{Irradiation as a form of dilution}

When suitable dye molecules absorb a photon they bend (photoisomerize) and
thereby weaken the ordering potentials causing the underlying nematic order, the tilt and the hindered
rotation about the molecular axis that ultimately gives the electric polarization in SmC* liquid crystals.
The first two orderings strengthen the latter.
One might imagine that for the soft interactions, giving thermal effects (in
contrast to the athermal effects of hard, steric packing), there are effective potentials and hence
probabilities of the form
\be
p(\psi) \propto \exp \left (- \frac{J f(\psi)}{k_B T} \right ),\label{eq:prob}
\ee
where $f(\psi)$ is a function of some suitable generalized coordinate $\psi$, and $J$ is a scale to the coupling
interaction. We further assume that $J$ scales according to $J=J_0 (1 - \delta + n\s{t}\delta)$, which suggests
that as neighboring molecules in a coordination shell are bent, $n\s{t} < 1$, they
contribute less or not at all to the ordering; here $J_0$ stands for interaction strength in the dark.
The fraction of molecules present that are photo-active is $\delta$. Here
$\delta \sim 0.05$ in the guest-host systems of Langhoff and
Giesselmann~\cite{giesselmann:03,giesselmann:02} and $\delta = 1$ in
the pure systems of SmC* photo-rods~\cite{giesselmann:04}. The fraction
$\delta$ would perhaps be smaller than its purely number fraction value because dye molecules
converted from \textit{trans} to \textit{cis} state still contribute to some extent to the ordering
(residual \textit{cis}-effect).
Rearranging via $n\s{t} = 1 - n\s{c}$, we obtain $J=J_0(1- \delta + n\s{t} \delta) =
J_0 (1 - n\s{c} \delta)$.
Overall, we can write the coupling strength relative to thermal effects in
Eq.~(\ref{eq:prob}) as
\be
\frac{J_0}{k\s{B} T/(1- n\s{c}(x) \delta)} \quad \Rightarrow \quad T \rightarrow
\frac{T}{1 - n\s{c}(x) \delta}.
\ee
The latter is the effective rise in temperature mapping to a particular
illumination, $\mathcal{I}(x)$, that determines $n\s{c}(x)$.

\subsection{Interpretation of measurements of non-uniform $P\s{s}(x)$}

Commonly, spontaneous polarization is measured by applying a triangular voltage~\cite{triangular:01}
across a cell filled with SmC* liquid crystal; see Fig.~\ref{fig2}. On reversing this voltage, the polarization also reverses
leading to a flow of polarization charge.
We leave aside the question whether the system in
the field-free state is helical, surface stabilized~\cite{clarklagerwall:80} or in the
\textbf{V}-switching state with $\bm{P}\s{s}$ in the ($x$,$y$) plane; see
\v{C}opi\v{c} \textit{et al.}~\cite{copic:02} for the underlying physics.
Langhoff and Giesselmann do not specify which their systems are,
but they have samples of thickness $L = 2$ $\mu$m so a helical ground state is unlikely.

Taking into account that \textit{cis} concentration varies with depth, it is
clear that spontaneous polarization should be non-uniform.
The polarization perceived to be switched is
\bea
\overline{P\s{s}} (T,I_0) &=& \frac{1}{L} \int_0^L {\rm d}x P\s{s}(T,x)
\nonumber \\
&=& \frac{1}{L} \int_0^L {\rm d}x P\s{s}^0 \left ( \frac{T}{1- n\s{c}(x)
\delta } \right ),
\label{meanps}
\eea
where $P\s{s}(T,x)$ is the actual polarization at depth $x$.
By contrast $P\s{s}^0 (T/(1- n\s{c} \delta))$ is our
model of polarization at $x$, namely the dark-state polarization but
shifted to higher effective temperatures if $n\s{c}(x) > 0$.
We apply our model of dilution simply to polarization, leaving aside the
question of tilt angle.

Ideally the aim of experiment is to measure photo-induced depression of
the dark state polarization $P\s{s}^0(T)$ to the value $P\s{s}(T,I_0)$
characteristic of the incident intensity.  However, in a real
measurement, the depression to $P\s{s}(T,I_0)$ only takes place at the
incidence surface, and is less deeper in the sample since the intensity
drops; $I(x) < I_0$.  Being a depth average, $\overline{P\s{s}} (T,I_0)$
is depressed less than the depression to $P\s{s}(T,I_0)$.  Thus the
measured change is an underestimate of that suffered by a system
uniformly exposed to $I_0$.

\subsection{Comparison with experiment}

\begin{figure}
  \includegraphics[width=9.4cm]{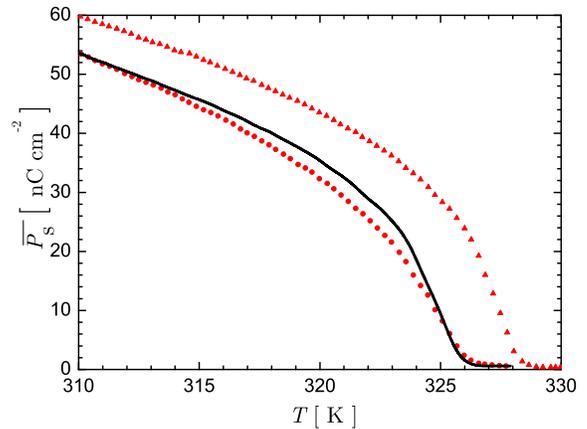}
  \caption{Spontaneous polarization as a function of temperature calculated by
using relation (\ref{meanps}), as explained in the main text, for the values of
fitting parameters $\delta=0.019$, $\tau_0 = 27.76 \times 10^{-12}$ s and $\kappa=
9765$ K (solid line). For comparison, we have shown experimental results
in the irradiated state (circles). Experimental results for the dark state (triangles), which we
used as an input to obtain our curve, are also presented.}
  \label{fig3}
\end{figure}

To test the above modeling strategy, we take Langhoff and Giesselmann's~\cite{giesselmann:03}
dark-state $P\s{s}^0(T)$ and match it to photo-modifications they obtain. Since a field is applied
to measure polarization by its reversal, the dark-state $P\s{s}^0(T)$ clearly has a foot for $T>T\s{AC}$ (due to the
electroclinic effect where tilt is induced by a field~\cite{garoffmeyer:77}) and is also modified for $T<T\s{AC}$ away
from any singular, zero field result (see Fig.~\ref{fig3}).

At finite incident illumination, $I_0$, the fit to the experimental $\overline{P\s{s}}(T,I_0)$
of Eq.~(\ref{meanps}) is good (Fig.~\ref{fig3}). Since the characteristics of the absorption,
$d\s{B}$ and $\alpha$, are not available, we have estimated $d\s{B} = 3.2$ $\mu$m from the known absorption of
azobenzene~\cite{vanOosten:01}, but adjusted for the different molecular number densities of azobenzene
and the dye used by Langhoff and Giesselmann.
Some factors entering the parameter $\alpha = \eta I_0 \Gamma \tau = I_0 \eta \tau / \gamma d\s{B}$ are
taken from experiment ($I_0= 75$ $\mu$W cm${}^{-2}$ and $\lambda=365$ nm), while some of them can be estimated ($\eta$ and $\gamma$).
For instance the quantum efficiency can be taken~\cite{ikedabook:09} as $\eta \simeq 0.6$ in these azo-based dyes
and $\gamma = \hbar \omega n\s{d} = 5.4 \times 10^7$ J m${}^{-3}$. Still, we have two free parameters, $\tau_0$ and $\kappa$, entering $\alpha$ through $\tau = \tau_0 \e^{\kappa/T}$. In addition, we use $\delta$ as a third fitting parameter. This parameter
is a measure of how much ordering influence of a dye molecule is lost when it is transformed to the \textit{cis}
state, and is therefore expected to be different from the bare dye number fraction $5\%$. Thus three fitting parameters
remain at our disposition. The best overall fit presented in Fig.~\ref{fig3}, we obtained with the following
values of fit parameters: $\delta = 0.019$, $\tau_0 = 27.76 \times 10^{-12}$ s and $\kappa = 9765$ K. In our analysis, the
parameter $\alpha(T)$ changes in the range ($0.67$, $3.46$) which corresponds to the nonlinear (i.e. non-Beer)
absorption regime. The departure of $\overline{P\s{s}}(T,I_0)$ from the data at lower temperatures
is perhaps a measure of the athermal (packing) effects mentioned above. Future experiments could measure
absorption characteristics $d\s{B}$ and $\alpha(T)$ and thus reduce the number of fit parameters to only one -- our $\delta$.

\subsection{Analysis of pure dye systems}

Examination of the experimental $\overline{P\s{s}}(T,I_0) - T$ curves shows
that for a given $I_0$ there are temperatures at which the polarization is uniform through the thickness, namely
$P\s{s}(T,x)=0$ since the average $\overline{P\s{s}}$ that is measured vanishes. At all lower temperatures the
polarization profile must be non-uniform.
One can draw a quite powerful conclusion from this observation. In the experiments we quoted in
the previous section the number fraction
of dye molecules was $5\%$. Given that there is still a profile of light intensity and hence also of $n\s{c}(x)$, even
when $P\s{s}(T,x)=0$ for all $x$ between $0$ and $L$, then $P\s{s}$ must be vanishing at the back face when $n\s{c}<1$
and hence an ''impurity'' level of less than $5\%$ is sufficient to eliminate $P\s{s}$.

The low level of required molecular transformation to achieve the maximal polarization change suggests a possible explanation of the perplexing results of Saipa \textit{et al.}~\cite{giesselmann:04}.
These authors found that the greatest photo-induced changes of polarization
occurred for irradiation at a wavelength $\lambda = 450$ nm, near the \textit{cis} absorption maximum, rather than at
$\lambda \simeq 360$ nm, the \textit{trans} absorption maximum where conversion is most likely and naively polarization
reduction would be most efficient.
Their system was composed $100\%$ of the dye W470 with a huge dark state polarization of up to $230$ nC cm${^{-2}}$.
In a $100\%$ dye system, irradiated with light of wavelength $\lambda \simeq 360$ nm, the Beer length is short, presumably $d\s{B} \ll L$.
Then at low intensities conversion to \textit{cis} is in a thin layer $\sim d\s{B}$.  This localization of conversion
means that $\overline{P\s{s}} \sim (1- d\s{B}/L) P\s{s}^0$, that is the change in average polarization is very little. To convert over a greater depth, and therefore to get a greater depression of $\overline{P\s{s}}$, one has to increase the incident intensity, that is,
$\alpha$. Then the conversion in the initial layers must be to a significant $n\s{c} \lesssim 1$ in order to eliminate
absorbers (\textit{trans} isomers) from the path. Then the average $\overline{P\s{s}} \rightarrow 0$ since conversion deeper down can take place. However, this strategy is very inefficient since significant volumes of the
sample have $n\s{c} \sim 1$ rather than $n\s{c} \sim 0.05$ or less that we have
seen is already adequate to obtain complete loss of $P\s{s}$. Far more
effective in this experiment was to have light well de-tuned from the \textit{trans} to
\textit{cis} absorption maximum. Thus, conversion was lower, but perhaps sufficient to achieve
$P\s{s}(T,x)=0$, and light being de-tuned penetration was deeper. Then there would be a
greater volume with $P\s{s} = 0$, and thus a greater change in $\overline{P\s{s}}$.
However, this is an inefficient way to proceed since much \textit{cis} absorption of photons
also takes place.

\section{Dynamics}\label{sect:dyn}

Since the depression of polarization rests upon the number fraction $n\s{c}$ of
\textit{cis} molecules, the dynamics of $\overline{P\s{s}}(T,I_0,t)$ is specified by the dynamics of
$n\s{c}(x,t)$ which we now sketch.
\begin{figure}
  \includegraphics[width=9.4cm]{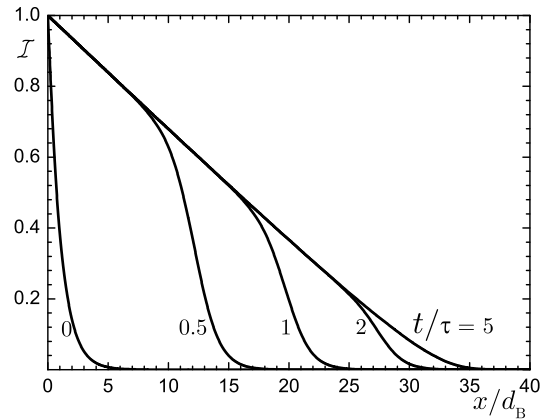}
  \caption{Reduced intensity $\mathcal{I}=I/I_0$ as a function of reduced depth
$x/d\s{B}$ for $\alpha=30$
  and different reduced times $t/\tau=0$, $0.5$, $1$, $2$ and $5$.}
  \label{fig4}
\end{figure}

Beer absorption strictly has no dynamics since it holds only if the number
fraction $n\s{t}$ is unchanging; $\partial \mathcal{I}/\partial x = - \mathcal{I}/d\s{B}$. Equally, there could
be no change in $P\s{s}$ since $n\s{c}=0$ in this limit. To explore the dynamics of non-Beer absorption, we need to solve
the coupled Eqs.~(\ref{intensityr}) and (\ref{transnumber}).
It can be shown~\cite{corbettdyn:01} that these simultaneous, non-linear
partial differential equations can be reduced to a quadrature
\be
\frac{t}{\tau} = \int_{\mathcal{A}}^{x/d\s{B}} \frac{{\rm d}
\mathcal{A}'}{\alpha + \mathcal{A}' - x/d\s{B} - \alpha
\e^{-\mathcal{A}'}},
\label{quadrature}
\ee
where $\mathcal{A} = - \ln (\mathcal{I})$ is the absorption (here logs base e). At $t=0$, before
any conversion has had time to deplete the (\textit{trans}) absorbers, $n\s{t}=1$ and
$\partial \mathcal{I}/\partial x = - \mathcal{I}/d\s{B}$ simply; the $\mathcal{I}(x,t=0)$ profile is of the
Beer form and $\mathcal{A}\s{B} = x/d\s{B}$. The lower limit tends to the upper causing the integral to vanish and thus $t/\tau
= 0$, as required. For $\mathcal{A} \rightarrow \mathcal{A}\s{eq}$, the equilibrium nonlinear (non-Beer) absorbance, the integral diverges since Eq.~(\ref{eqsolution}) for $\mathcal{A}\s{eq}$ is nothing other
than the vanishing of the denominator in Eq.~(\ref{quadrature}) -- hence $t/\tau \rightarrow \infty$.
From the solution $\mathcal{A}(x/d\s{B},t/\tau)$ of (\ref{quadrature}) we obtain $\mathcal{I} = \e^{-\mathcal{A}}$.
This nonlinear absorption dynamics has been thoroughly verified by the
experiments of Serra and Terentjev~\cite{serra:08}.
As time increases, and for intense enough beams ($\alpha \gtrsim 1$),
depletion of $n\s{t}(x,t)$ means greater penetration of $\mathcal{I}$. In effect a front of conversion
penetrates the sample giving a growing volume where $P\s{s}(T,x,t)$ is depressed; see Fig.~\ref{fig4}.

\begin{figure}
  \includegraphics[width=9.4cm]{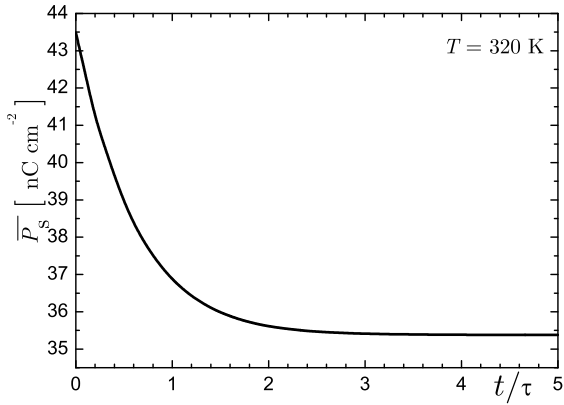}
  \caption{Spontaneous polarization (\ref{meanpsdyn}) as a function of reduced
time $t/\tau$ for $T=320$ K
  and the values of $\delta$, $\tau_0$ and $\kappa$ as in Fig.~\ref{fig3}.}
  \label{fig5}
\end{figure}

Armed with $\mathcal{A}(x,t)$, we have $n\s{t} = d\s{B} \partial
\mathcal{A}/\partial x$ on rearranging (\ref{intensityr}),
and hence $n\s{c} = 1 - n\s{t}$ which generalizes Eq.~(\ref{meanps}) to
the time-dependent polarization
\be
\overline{P\s{s}} (T,I_0,t) = \frac{1}{L} \int_0^L {\rm d}x P\s{s}^0
\left ( \frac{T}{1- n\s{c}(x,t) \delta } \right ).
\label{meanpsdyn}
\ee
Fig.~\ref{fig5} shows how the measured polarization is depressed by illumination only as a front of conversion
to \textit{cis} advances through the sample thickness.

The switch-on dynamics generally takes place over a period $t \lesssim \tau$,
depending on the incident intensity, $I_0$,
that is on $\alpha$, and on the thickness $L/d\s{B}$ for the conversion front
to propagate through; see Fig.~2 of [\onlinecite{corbettdyn:01}] where this is
demonstrated for a range of thicknesses and incident light intensities. The
characteristic on-time approaches $\tau$ for thicker samples where $L/d\s{B} \sim \alpha$.

Switch-off dynamics is driven by the response $\dot{n}\s{t}$ of
(\ref{transnumber}) when $I=0$, that is
\be
n\s{c} (x,t) = n\s{c}(x,0) \e^{-t/\tau},
\label{switchoff}
\ee
where now $t$ is the time since switch-off. The $n\s{c}(x,0)$ number fraction
at that time could either be the equilibrium value $n\s{c}(x) = \alpha \mathcal{I}(x)/(1+ \alpha
\mathcal{I}(x))$ or simply the value attained by $t=0$ when the switch-on was concluded.
After inserting (\ref{switchoff}) into $\mathcal{A}(L,t) = (1/d\s{B}) \int_0^L {\rm d}x \, (1- n\s{c}(x,t))$,
one obtains
\be
\mathcal{A}(L,t) = \frac{L}{d\s{B}} - \left ( \frac{L}{d\s{B}} - \mathcal{A}(L,0) \right ) \e^{-t/\tau},
\label{absrec}
\ee
where $\mathcal{A}(L,0)$ is the absorption at $t=0$.
Experiments of Langhoff and Giesselmann~\cite{giesselmann:02} revealed that both absorption and spontaneous polarization
increase following monoexponential laws with the same characteristic times. In view of our relation (\ref{absrec}) it is clear
that this characteristic time for absorption should be equal to the thermal relaxation time $\tau$.
We have found, however, that a simple monoexponential law for the spontaneous polarization breaks down in the vicinity of the critical point, while it holds outside the critical region.

Given $n\s{c}(x,t)$ is deeply buried in (\ref{meanpsdyn}) the off-dynamics is
not necessarily simple. However we can expand $P\s{s}^0$ in (\ref{meanpsdyn}) assuming that $n\s{c}(x,t)
\delta$ is small
\bea
&\overline{P\s{s}}&(T,I_0,t) = P\s{s}^0 + T \frac{\partial
P\s{s}^0}{\partial T} \frac{\delta}{L}
\int_0^L {\rm d} x \, n\s{c}(x,t) + \dots \nonumber \\
&=& P\s{s}^0 + T \frac{\partial P\s{s}^0}{\partial T}
\frac{\delta}{L} \, \e^{-t/\tau} \int_0^L {\rm d} x \, n\s{c}(x,0) + \dots \nonumber \\
&=& P\s{s}^0 + T  \frac{\partial P\s{s}^0}{\partial T}
\left ( 1 - \frac{d\s{B}}{L} \mathcal{A}^0 \right ) \e^{-t/\tau} \delta
+ O(\e^{-2t/\tau})
\eea
where $\mathcal{A}^0 = \mathcal{A}(L,0)$.
\begin{figure}
  \includegraphics[width=9.4cm]{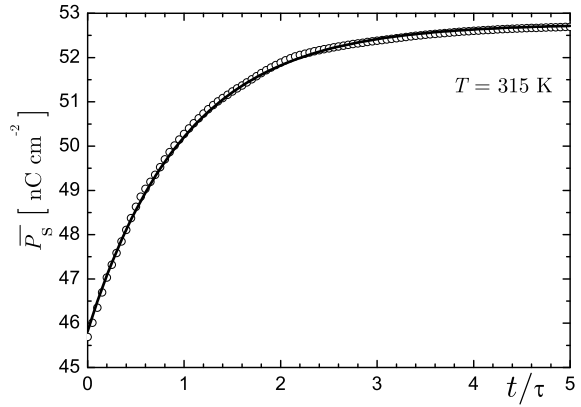}
  \caption{Polarization versus reduced time for $T=315$ K and the values of $\delta$, $\tau_0$ and $\kappa$ as in Fig.~\ref{fig3}.
  The ''exact'' data, i.e. those obtained by using the integration procedure indicated by formula (\ref{meanpsdyn}) are presented
  by circles, while the solid curve represents the fit to the monoexponential curve $A+B\e^{-t/\tau}$, where $\tau$ is the thermal
  relaxation time $\tau$ and $A$ and $B$ are fit parameters.}
  \label{fig6}
\end{figure}

\begin{figure}
  \includegraphics[width=9.4cm]{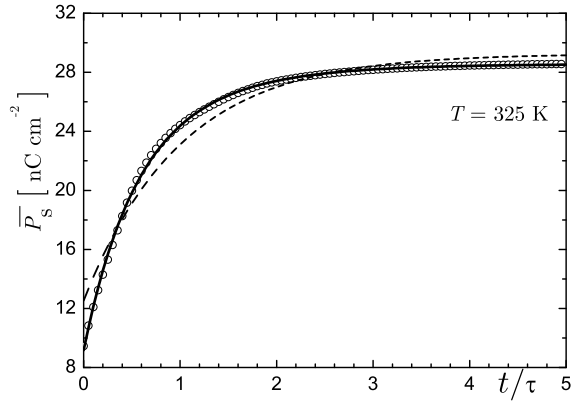}
  \caption{The same as in Fig.~\ref{fig6}, but for $T=325$ K, closer to the critical temperature $T\s{AC} \approx 327$ K.
  The data obtained by using formula (\ref{meanpsdyn}) are presented by circles, monoexponential fit by the dashed line, and
  biexponential fit of the form $A+B\e^{-t/\tau}+C\e^{-2t/\tau}$ by the solid line.}
  \label{fig7}
\end{figure}

We have shown that $\overline{P\s{s}}(T,I_0,t)$ has an expansion in $\e^{-t/\tau}$.
As more terms in the expansion are required to describe
$P\s{s}^0(T/(1-n\s{c}\delta))$ more precisely, then so are faster
decay terms $\e^{-2t/\tau}$, ... represented.
Unfortunately, coefficients of this expansion are not available from existing experiments, so we tried to fit
time-dependent polarization with the simple form $\overline{P\s{s}} = A+B\e^{-t/\tau}+C\e^{-2t/\tau}$,
with $A$, $B$ and $C$ being the fit parameters. One can see that the monoexponential form ($C=0$)
works fairly well outside of the critical region (Fig.~\ref{fig6}), but not so well in the vicinity
of the critical point (the dashed line in Fig.~\ref{fig7}). If one uses the biexponential form, $C \neq 0$,
the fit becomes much better (Fig.~\ref{fig7}).

\section{Summary and outlook}

We have argued, given light is absorbed in driving the photoferroelectric
effect, that the mechanism and details of absorption are crucial to understanding the optical depression
of electric polarization. To achieve a polarization response to illumination, the ground state population
of dye molecules has to be materially changed. From this it follows that the intensity profile is
non-Beer (non-exponential). Equally there must be a characteristic dynamics for the switch-on phase where a
front of dye depletion progresses through the thickness. Non-uniformity of the polarization means that
measurements of the depression must underestimate depression assuming a uniform intensity $I_0$ through the sample.
This underestimate has implications for modeling the balance of primary
and secondary influences in the photoferroelectric effect.

Future experiments should simultaneously measure the absorption in detail,
namely the Beer length $d\s{B}$ for weak illumination and the nonlinear absorption in more intense cases to yield
the effective incident intensity (our quantity $\alpha$). Thereby only one parameter would remain, our
$\delta$, which is a measure of how much the ordering effect of a dye molecule is lost when it is bent (on
entering the \textit{cis} state).

\section*{Acknowledgements}

MK thanks the Winton Programme for the Physics of Sustainability and the Cambridge Overseas Trust for financial support, and MW thanks the EPSRC for a Senior Fellowship.

\bibliography{Knezevic}
\end{document}